\documentclass[pra,floatfix,preprint,superscriptaddress,aps,longbibliography]{revtex4-1} 

\usepackage{graphicx}
\usepackage{textcomp}
\usepackage[english]{babel} 
\usepackage[utf8]{inputenc}
\usepackage[T1]{fontenc}
\usepackage{hyperref}
\usepackage{amssymb}
\usepackage{amsmath}
\usepackage[amssymb]{SIunits}
\usepackage{color}
\usepackage[version=3]{mhchem}
\usepackage{lineno}

\newcommand{\CNRSAddress}{Univ. Grenoble Alpes, CNRS, Institut Néel, "Nanophysique et semiconducteurs" group, F-38000 Grenoble, France}
\newcommand{\CEAAddress}{Univ. Grenoble Alpes, CEA, INAC, PHELIQS, "Nanophysique et semiconducteurs" group, F-38000 Grenoble, France}
\newcommand{\DTUAddress}{DTU Fotonik, Department of Photonics Engineering, Technical University of Denmark, Ørsteds Plads, Building 343, DK-2800 Kongens Lyngby, Denmark}

\begin{document}
\title{Enhanced photon extraction from a nanowire quantum dot using a bottom-up photonic shell}

\author{Mathieu Jeannin}
\thanks{contributed equally to this work}
\affiliation{\CNRSAddress}

\author{Thibault Cremel}
\thanks{contributed equally to this work}
\affiliation{\CEAAddress}

\author{Teppo Häyrynen}

\author{Niels Gregersen}
\affiliation{\DTUAddress}

\author{Edith Bellet-Amalric}
\affiliation{\CEAAddress}

\author{Gilles Nogues}
\email{gilles.nogues@neel.cnrs.fr}
\affiliation{\CNRSAddress}

\author{Kuntheak Kheng}
\affiliation{\CEAAddress}

\begin{abstract}
Semiconductor nanowires offer the possibility to grow high quality quantum dot
heterostructures, and in particular CdSe quantum dots inserted in ZnSe nanowires
have demonstrated the ability to emit single photons up to room temperature. In
this letter, we demonstrate a bottom-up approach to fabricate a photonic
fiber-like structure around such nanowire quantum dots by depositing an oxide
shell using atomic layer deposition. Simulations suggest that the intensity collected in our
NA=0.6 microscope objective can be increased by a factor 7 with respect to the bare
nanowire case. Combining micro-photoluminescence, decay time measurements and
numerical simulations, we obtain a 4-fold increase in the collected photoluminescence from the quantum dot. 
We show that this improvement is due to an increase of
the quantum dot emission rate and a redirection of the emitted light. Our
ex-situ fabrication technique allows a precise and reproducible fabrication on a
large scale. Its improved extraction efficiency is compared to state of the art
top-down devices.
\end{abstract} 

\maketitle

\section{Introduction}

Controlling and enhancing the spontaneous emission of quantum emitters is one of
the current key issues in the field of nanophotonics. Semiconductor quantum dots
(QDs) are considered as promising and efficient single-photon emitters for
quantum optics applications.
\cite{Michler2000,Santori2001,Santori2002,Zrenner2002,Akopian2006,Shields2007}
Over the past few years, several approaches have been pursued to control their
emission properties, from the use of photonic crystals
\cite{Viasnoff-Schwoob2005,Lund-Hansen2008} to top-down photonic wires
\cite{Claudon2010,Heinrich2010,Bleuse2011} and
trumpets.\cite{Munsch2012,Munsch2013} These strategies are based on the early
work of Purcell\cite{Purcell1946} which demonstrated that the
spontaneous emission of an emitter can be modified by engineering its
electromagnetic environment. They rely on a waveguiding approach to increase the
coupling between a well-defined propagating optical mode and the QD while
simultaneously reducing the coupling between the QD and background radiation
modes, offering control of both the optical mode profile and the QD spontaneous
emission rate.

In this context, the interest of the dot-in-a-nanowire configuration fabricated
using bottom-up methods naturally arises because it provides a simple way to
ensure the centering of a single quantum emitter in the photonic
structure.\cite{Reimer2012,Bulgarini2012,Bulgarini2014} The bottom-up
fabrication method also avoids heavy processing, like etching the semiconducting
material, that is often detrimental to the QDs optical properties. However, the
main realizations up to now concern III-V semiconductors,
\cite{Reimer2012,Bulgarini2012,Bulgarini2014}  limiting the operation range to
the cryogenic temperature.
Tackling this issue, the potential of II-VI materials, in particular CdSe QDs
inserted inside ZnSe nanowires (NWs) has been demonstrated in previous studies.
They allow for robust high temperature single-photon emission using
heteroepitaxial \cite{Tribu2008} or homoepitaxial \cite{Bounouar2012a} nanowire
growth. Contrary to all the aforementioned systems where the photonic wire
structure has a diameter comparable to the wavelength $\lambda/n$ of the guided
light which allows for highly efficient coupling to the HE$_{11}$
mode\cite{Nowicki2008}, the diameter of the II-VI NW embedding the QD
($\sim$\unit{20}{\nano\meter}) is much smaller than the wavelength of the
emitted light (\unit{530}{\nano\meter}). It leads to light emission
predominantly into non-guided radiation modes and a low collection efficiency.
An additional fabrication effort has thus to be made to ensure an efficient
coupling to the collection optics.

In a previous report\cite{Cremel2014} we have theoretically investigated the
potential of using an oxide shell deposition on a bare ZnSe NW to form a thick
photonic wire structure. In this article, we experimentally demonstrate the use
of atomic layer deposition (ALD) to fabricate a conformal aluminum oxide
(Al$_2$O$_3$) shell around ZnSe NWs containing a single CdSe QD. We show that
the oxide shell drastically enhances the light intensity emitted by the QD, and
we use time-resolved microphotoluminescence to systematically study the effect
of the shell thickness on the nanowire quantum dot (NWQD) emission rate. Our
results are compared to numerical simulations accounting for the real NW
geometry, evidencing the different physical mechanisms leading to the
enhancement of the spontaneous emission from the QD and to the improved light
collection from the emitting structure.

\section{Principles of operation}

\begin{figure}
	\centering
	\includegraphics[width=8.6cm]{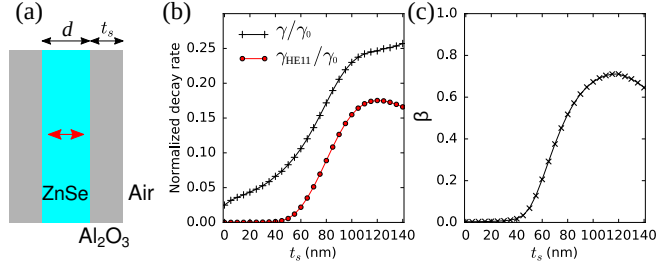}
	\caption{(a)  Geometry of the infinite NW. (b) Total spontaneous emission rate (black $+$) and spontaneous emission rate into the first guided mode HE$_{11}$ (red \textbullet) as a function of shell radius for a radial dipole. (c) Fraction $\beta$ of power radiated into the HE$_{11}$ mode.
	}
	\label{fig:Semianalytical}
\end{figure}

To illustrate the effect of the NW and  its surrounding medium
on the QD emission rate, let us consider a QD placed inside an
infinitely long cylinder as illustrated in Fig.~\ref{fig:Semianalytical}(a)
radiating a field at a wavelength $\lambda$. The cylinder is made of a
dielectric material (refractive index $n$) and has a diameter $d$. We first
consider a dipole orientation perpendicular to the NW axis in order to use the
NW as a propagation medium for the emitted light. In the limit where $d \ll
\lambda/n$, the dielectric screening effect\cite{Bleuse2011} reduces the
spontaneous emission rate $\gamma$  by a factor:
\begin{equation}
\frac{\gamma}{\gamma_0}=\frac{4}{n(n^2+1)^2},\label{eq:screening}
\end{equation}
where $\gamma_0$  is the radiative emission rate in the bulk
material of index $n$.\cite{ClaudonGerard_HarnessingLightwith_13}  For a ZnSe cylinder
($n_{\rm ZnSe}$  = 2.68 at $\lambda$=\unit{530}{\nano\meter}), the screening
factor is $\sim$1/45. If the NW is surrounded by a shell of refractive index
$n_s$ instead of vacuum, equation \ref{eq:screening} remains valid by replacing
$n$ with the index contrast $n/n_s$. For an \ce{Al2O3} surrounding medium
($n_s=$1.77), the screening factor becomes $\sim$1/4.1, resulting in an order
of magnitude larger radiative rate.

In addition to changing the dielectric screening, the \ce{Al2O3} shell also
influences the guiding of light along the NW. We have computed the total
emission  rate $\gamma$ and the emission rate $\gamma_{\rm HE11}$ into the
fundamental HE$_{11}$ waveguide mode from a radial dipole as function of the
shell thickness $t_s$ [see Fig.~\ref{fig:Semianalytical}(a)] using a
semi-analytical approach\cite{Yariv1997} combined with an efficient non-uniform
discretization scheme in k
space.\cite{HaeyrynenGregersen_OpengeometryFourier_16} The results are plotted
in Figure \ref{fig:Semianalytical}(b). We observe that the shell thickness of
$\sim$\unit{120}{\nano\meter} not only leads to an increased total emission
rate, it also allows for confinement of the fundamental HE$_{11}$ mode to the
core-shell NW leading to a preferential coupling of the emitted light to this
mode. Figure \ref{fig:Semianalytical}(c) presents the spontaneous emission
$\beta$ factor representing the fraction $\beta=\gamma_{\rm HE11}/\gamma$  of
emitted light coupled to the HE$_{11}$ mode. We observe indeed that up to 71\%
of the emitted light is coupled to this mode for $t_s$=\unit{120}{\nano\meter}.
The dipole thus becomes coupled to the equivalent of a monomode photonic
wire\cite{Claudon2010,Heinrich2010,Bleuse2011,Reimer2012,Bulgarini2012,Bulgarini2014} 
paving the way to the control of its far-field radiation pattern.

\section{Sample fabrication}

\begin{figure}
	\centering
	\includegraphics[width=7cm]{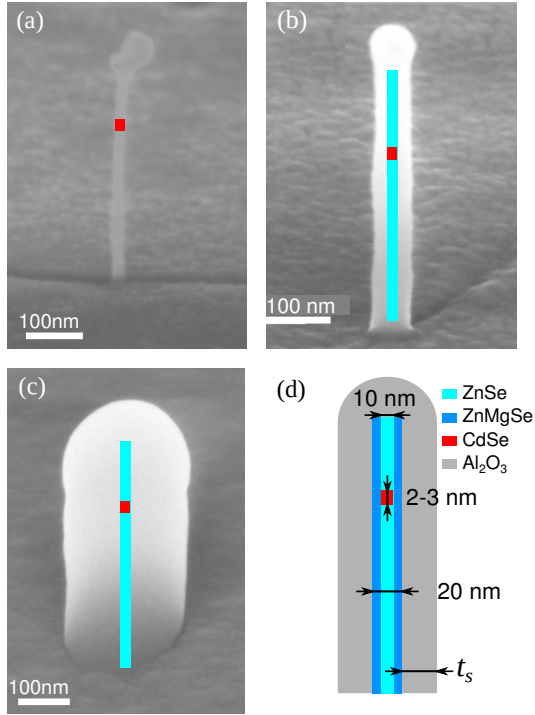}
	\caption{
		(a) SEM image of a standing ZnSe/ZnMgSe NW embedding a CdSe QD. The QD
		position is marked by the red square. (b,c) Tilted SEM image of a ZnSe
		NW after a \unit{20}{\nano\meter} and \unit{110}{\nano\meter} thick
		Al$_2$O$_3$ shell deposition respectively. The NW is sketched on the SEM
		image. Note the circular shape of the shell as well as its hemispherical
		termination above the NW apex. (d) Sketch of the NWQD geometry,
		indicating the QD height (2-\unit{3}{\nano\meter}), the NW diameter
		($\simeq$\unit{10}{\nano\meter}), the epitaxial shell thickness
		($\simeq$\unit{5}{\nano\meter}) and the ALD shell thickness $t_s$.
	}
	\label{fig:NWPresentation}
\end{figure}

Our emitters are CdSe QDs embedded inside a ZnSe NW with a thin, epitaxial
passivation Zn$_{0.83}$Mg$_{0.17}$Se shell grown around the NW. They are grown
by molecular beam epitaxy on a GaAs(111)B substrate. A ZnSe buffer layer is
first grown on the GaAs substrate after which a thin layer of Au (less than one
monolayer thick) is evaporated on the sample surface and dewetted at
\unit{510}{\celsius} to form small ($\sim$\unit{10}{\nano\meter} diameter) Au
droplets that serve as a catalyst for the NW growth. The substrate temperature
is then set at \unit{400}{\celsius} and a flux of Zn and Se atoms with an excess
of Se is used, inducing preferential growth of vertical ZnSe NWs. The NWs are in
wurtzite phase and their diameter is the same as the droplet
($\sim$\unit{10}{\nano\meter} diameter). The thickness of the initial Au layer
is chosen to ensure a low NW density ($\leq$ 1 NW per
\unit{}{\micro\meter\squared}). After the growth of a \unit{400}{\nano\meter}
high NW, the atom fluxes are stopped to allow the evacuation of residual Se
atoms inside the droplet. Then, the QD is grown under a flux of Cd and Se atoms
for \unit{20}{\second}. The fluxes are interrupted again before the ZnSe growth
is resumed, resulting in an expected QD height of 2-\unit{3}{\nano\meter}
inserted in a $\sim$\unit{700}{\nano\meter} high NW. Finally, an epitaxial
Zn$_{0.83}$Mg$_{0.17}$Se shell (\unit{5}{\nano\meter} thick) is grown around the
NW at \unit{220}{\celsius}. A scanning electron microscope (SEM) image of such a
CdSe/ZnSe/ZnMgSe core/shell NWQD system is presented in
Figure~\ref{fig:NWPresentation}(a). The flag-shape termination of the NW is formed
during the growth of the ZnMgSe shell. It is present in some NWs.

The higher bandgap of Zn$_{0.83}$Mg$_{0.17}$Se shell prevents the charge
carriers to recombine non-radiatively on the ZnSe NW sidewall and hence improves
the quantum yield of the CdSe emitter. In principle, it could directly be used
to grow a photonic wire of diameter $\sim \lambda / n_{\rm ZnSe}$ around the
NWQD. However, during the epitaxial shell growth two phenomena are competing:
the radial growth of the shell around the wurtzite NWs, and the vertical growth
of a 2D Zn$_{0.83}$Mg$_{0.17}$Se layer on the sample surface. The radial shell
growth rate is very low because the growth of ZnSe on WZ surfaces is not
favourable. Because of this low shell growth rate, a trade-off has to be found
to avoid burying the NWs in a Zn$_{0.83}$Mg$_{0.17}$Se matrix. As a result, only
thin epitaxial shells can be fabricated.

The complexity of creating a thick epitaxial shell is one of the reasons why we
fabricate the photonic structure by depositing an oxide shell around the NW
using ALD. Another reason is that, since this process step can be done
\emph{separately} from the NW growth process, it allows to tune ex situ the
shell parameters after a first optical characterization of the QD. Indeed, due
to its slow deposition rate, the ALD process allows to precisely control the
deposited thickness, which can also be finally verified using scanning electron
microscopy. We have tested several oxide materials, and selected Al$_2$O$_3$
because it produced very smooth and conformal, amorphous shells.
Figure~\ref{fig:NWPresentation}(b) and (c) show two SEM images of the resulting
oxide shell deposition (\unit{20}{\nano\meter} and \unit{110}{\nano\meter}), and
the complete structure is sketched in Fig.~\ref{fig:NWPresentation}(d). We note
that the conformal deposition allows to end the NW+shell structure by an almost
perfect half-sphere as can be seen in Fig.~\ref{fig:NWPresentation}(b,c). ALD
also buries the Au droplet under the shell. The latter might interact with the
field emitted by the QD through its localized plasmon resonance. Considering its
small diameter it will essentially absorb the incoming field. Moreover the
guided HE11 mode profile presents a minimum on the NW axis. This is why we
neglect the droplet influence in the following.

\section{Experimental results}

\begin{figure}
\centering
\includegraphics[width=8cm]{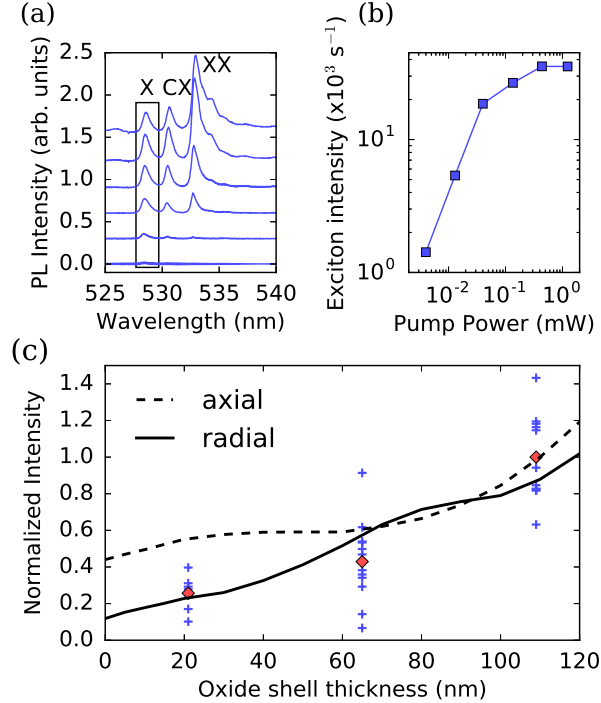}
\caption{
(a) \textmu PL spectrum of a NWQD with \unit{120}{\nano\meter} thick photonic
shell for different pumping powers. It shows a exciton (X), charged exciton (CX)
and biexciton (XX) lines. The corresponding pumping powers are reported in panel
(b). The black rectangle indicates the integration bandwidth used to extract the
total exciton emission intensity (X line). (b) Integrated exciton emission
intensity as a function of pumping power, in a log-log scale. (c) Blue crosses:
Total exciton emission intensity for different NWQDs as a function of the oxide
shell radius. Red diamonds: average of the experimental data points. Data are
normalized to the average intensity at $t_s=$\unit{110}{\nano\meter} Black
lines: results of the numerical simulations for a radial (solid line) and an
axial (dashed line) dipole. They are normalized to the axial intensity at
$t_s=$\unit{110}{\nano\meter}
}
\label{fig:IVsThick}
\end{figure}

A sample from a single epitaxial growth process is cut in pieces, and photonic
structures with different oxide shell thicknesses are fabricated. Taking
advantage of the low NW density, individual structures are optically
characterized directly on the growth substrate. The samples are mounted on the
cold finger of a He-flux cryostat and cooled down to \unit{4}{\kelvin}.
Individual photonic structures are probed using confocal microphotoluminescence
(\textmu PL).  They are excitated by a supercontinuum pulsed laser (Fianium
WhiteLase, \unit{10}{\pico\second} pulse duration, repetiton rate
\unit{76}{\mega\hertz}) and a spectrometer selecting a \unit{10}{\nano\meter}
bandwidth  centered around \unit{485}{\nano\meter}. This excitation energy,
below the ZnSe gap, allows us to induce crossed transitions between delocalized
states in the NW 1D continuum and a discrete confined 0D state in the NWQD band
structure\cite{VasanelliBastard_ContinuousAbsorptionBackground_02}. In this
configuration, the NW axis is aligned with the optical axis and emission from
the QD is collected by a $NA=0.6$ objective. A typical NWQD spectrum is
presented in Figure~\ref{fig:IVsThick}(a) as a function of the pump laser power.
Three lines can be identified and are attributed to the exciton (X), the charged
exciton (CX) and the bi-exciton (XX) respectively. The total emission intensity
of the X line as a function of the pump power is reported in
Figure~\ref{fig:IVsThick}(b). It shows a linear increase at low pumping power,
and a constant plateau at high pumping powers corresponding to the saturation of
the exciton level.\footnote{We note that single-photon emission is preserved if
	one integrates both the signal from the X and CX lines.\cite{Sallen2009}} Under
pulsed excitation, we note that changing the shell thickness might modify the
laser power in the NW and the excitation probability of the QD. Hence if affects
the slope at low power in Fig.~\ref{fig:IVsThick}(b). It has however no effect
on the saturation plateau which only depends on the QD emission rate and light
collection efficiency. This allows us to compare statistical sets of
nanostructures with different oxide shell thicknesses. The total integrated
emission at saturation as a function of the oxide shell thickness is reported in
blue markers for each NWQD in Figure~\ref{fig:IVsThick}(c). The values have been
normalized to the average intensity at $t_s=$\unit{110}{\nano\meter}. For NWs
without an oxide shell, the luminescence intensity is very low and we were never
able to reach the saturation regime, this is why we do not report the
corresponding points in Fig.~\ref{fig:IVsThick}(c). For each shell thickness, we
observe a large spread in exciton saturation intensity. However, we note a
general trend of increasing saturation intensity with increasing shell
thickness, as demonstrated by the red markers which show the position of the
average intensity of our measurements for each shell thickness. On average, the
deposition of a \unit{110}{\nano\meter} thick shell results in the experiments
in an almost 4-fold enhancement of the collected intensity with respect to the
\unit{20}{\nano\meter} thick shell case. The semi-analytical calculations show
that this enhancement is 10-fold when we compare to a NW without oxide shell.

\begin{figure}
	\centering
	\includegraphics[width=8.6cm]{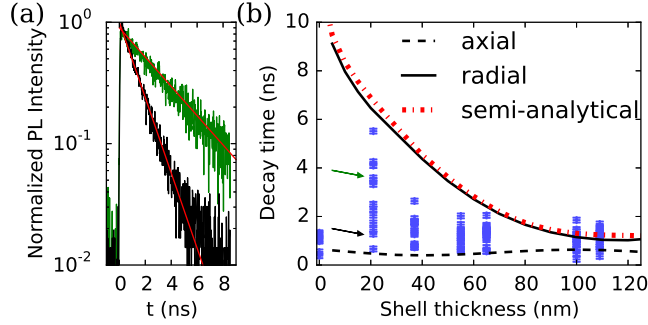}
	\caption{(a) Example of TRPL signal versus time for 2 NWs with a $t_s$=\unit{20}{\nano\meter} shell. Background counts are measured for $t<0$ and substracted. Amplitude of counts are normalized to 1 to compare the 2 datasets. Red lines are mono-exponential fits, whose corresponding points in (b) are showm by arrows. (b) Blue crosses: experimental exciton decay times for several QDs as a
	function of the oxide shell radius. The vertical error bars represent the fit
	error. Black lines: numerical simulation results
	for a radial dipole (solid line), or an axial dipole (dashed line). Red
	dashed-dotted line: Semi-analytical calculations for the infinite NW.
	}
	\label{fig:TRPLVsThick}
\end{figure}


The observed increase in intensity at saturation corresponds to the combination
of improved collection efficiency through light redirection from the structure
and enhancement of the spontaneous emission rate. In the latter case, a
modification of the QD dynamics is expected to be detected by measuring the
exciton decay rate.  Time-resolved measurements were carried out using a low
pump power as compared to the exciton saturation power to avoid any repopulation
of the X level. The measured decay transients are thus monoexponential. The
fitted decay constant is the total exciton decay time $\tau$. The experiment was
carried out in another setup on a different set of photonic structures compared
to the one of figure~\ref{fig:IVsThick}(c). The same excitation laser was used, 
the QD fluorescence was spectrally filtered in a spectrometer (500gr/mm grating) 
and integrated on an avalanche photodiode in a photon correlation setup, using 
the exit slit of the spectrometer as a spectral bandpass filter. 
The results of these measurements,
presented in Figure~\ref{fig:TRPLVsThick} show also a great dispersion in decay
time. One observes however that longer lifetimes are observed for smaller shell
thickness (up to \unit{5.9}{\nano\second}). Increasing the shell thickness leads
to an overall decrease in the measured exciton lifetime, hence an enhancement of
the exciton decay rate in agreement with the results of the numerical simulations.
For systems without an oxide shell, only a few NWQDs give a large enough signal to
be properly measured. They yield a much smaller dispersion of short decay times.

\section{Discussion and comparison to numerical simulations}

\subsection{Dispersion of the results}
For each oxide shell thickness, the large variations of the experimental results
in both Figs.~\ref{fig:IVsThick}(c) and \ref{fig:TRPLVsThick} have several
possible origins. First, the presence of non-radiative recombination channels
can reduce the intensity at saturation and change the decay time. The
non-radiative recombination rate can vary from QD to QD because of fabrication
inhomogeneities, leading to a spread in the measured values.\cite{Stepanov2015}
Second, variations in the QD aspect ratio and piezoelectric fields induced internal
strain applied by both the ZnSe core and the Zn$_{0.83}$Mg$_{0.17}$Se shell lead to
different overlap of electron and hole wavefunctions and hence different exciton
oscillator strengths. Finally, considering the QD aspect ratio and internal strain, we expect a heavy-hole exciton type for our
QDs.\cite{Eshelby1957,Eshelby1959,Zielinski2013,Ferrand2014} Heavy-hole exciton
recombination results in a mixture of circularly polarized emission, composed of
two degenerate out-of-phase radial dipoles. However, strain and confinement
effects might lead to valence band mixing between light hole and heavy hole
levels,\cite{Karlsson2006,Tonin2012,JeanninNogues_Lightholeexciton_17} resulting
in an emission composed of a mixture between axial and radial dipoles and hence
to a spread in total emitted intensity, as we discuss later. Additional
measurements on NWQDs grown in similar conditions and mechanically dispersed on a
substrate (i.e. lying horizontally on it) revealed that one NWQD out of 6 emit
light polarized along the NW axis, while others emit light polarized
perpendicularly to the NW axis, evidencing the presence of both kinds of
dipoles. Due to the Zn$_{0.83}$Mg$_{0.17}$Se shell and low temperature of
observation, we expect that non-radiative effects play a minor role. The epitaxial shell 
prevents non-radiative decay channels owing to surface traps. Additional measurements 
as a function of temperature show that both the emission intensity and the decay time 
do not change significantly up to 150-\unit{200}{\kelvin} (not presented here). 
This indicates that the non-radiative effects are not dominating at low temperature, 
as in the present experiment. 
While we cannot yet completely rule out the contribution of non-radiative effects, we 
think that the major effect to explain the dispersion of the results comes from 
variations in valence band mixing and oscillator strength due to the local environment of the QD. Finally let us stress that the shortest decay times (1-\unit{2}{\nano\second}) we measure remain longer than the decay time of CdSe self-assembled QD embedded in bulk ZnSe  (<\unit{1}{\nano\second})\cite{Bacher99}. The reduction of the dielectric screening effect is a main effect we evidence.

\subsection{Collected intensity and radiative lifetime}

To better understand the effect of the shell deposition on the NWQD emission, we
perform numerical simulations of the photonic structure formed by the full NW +
oxide shell geometry [see Fig.~\ref{fig:NWPresentation}(d)]. It takes into
account the presence of the ZnSe substrate, and the \ce{Al2O3} shell and layer
deposited on the NWs and substrate. The QD is modeled as an oscillating electric
dipole, either in the axial direction (along the NW axis) or in the radial
direction (orthogonal to the NW axis). We perform finite-element method
simulations (Comsol v4.1) to compute the total field radiated by the
dipole.\cite{JeanninNogues_Lightholeexciton_17} For each shell thickness and
dipole orientation, we evaluate the power radiated towards the objective by
computing the flux of the Poynting vector over a surface limited by its
numerical aperture (NA=0.6) in a region far from the NW where near field can be
neglected. The results of these simulations are reported in
Figure~\ref{fig:IVsThick}(c) in black lines for an axial (dashed line) or radial
(solid line) dipole. The results are normalized to the axial intensity at
$t_s=$\unit{110}{\nano\meter}. Comparing the simulated integrated intensity in
the case of a \unit{20}{\nano\meter} and \unit{110}{\nano\meter} reveals an
enhancement factor less than 2-fold for an axial dipole and almost 4-fold for a
radial dipole. The 4-fold enhancement observed in our measurements suggests that
on average, the dominant emitting dipole in our structure is radial, in good
agreement with the recombination of a heavy hole exciton.

The theoretical limits for the radiative lifetimes is extracted from the
numerical simulations by integrating the total power radiated over
every direction for the two dipole orientations (radial and axial) $P$. We
normalize this value by the same quantity computed for a dipole in bulk
\ce{ZnSe} $P_0$. For a purely radiative system we have $P/P_0 =
\gamma/\gamma_0=\tau_0/\tau$ \cite{Novotny2012}, where $\tau$ and $\tau_0$ are
the radiative lifetime for the nanostructure and for bulk \ce{ZnSe}
respectively. Radiative times are presented in black lines in
Figure~\ref{fig:TRPLVsThick}, where we have chosen
$\tau_0=\unit{300}{\pico\second}$ in good agreement with previously reported
radiative lifetime of CdSe QD in bulk
ZnSe\cite{FlissikowskiHenneberger_PhotonBeatsfrom_01}. The axial dipole radiates
with an almost constant decay time as a function of the oxide shell thickness,
while the radial dipole decay time strongly decreases with increasing oxide
shell thickness $t_s$. Additionally, we compare the decay time for the radial
dipole computed for the full geometry to the semi-analytical calculations for the
infinite NW presented in fig.~\ref{fig:Semianalytical}(b) with the same $\tau_0$
value. The agreement is excellent indicating that interference effects due to
reflections from the substrate and from the top hemispherical termination are
negligible.

Comparing the trends of the simulations, we can confirm that our emitters bear a
strong radial dipole character. The  measurements dispersion can be well
understood by considering that the real emitters are a mixture of radial and
axial dipoles radiating with a characteristic decay time comprised between the
simulated lifetimes of the pure radial and axial dipole. 
We do not observe long decay time for  NWQDs without an oxide shell in
Fig.~\ref{fig:TRPLVsThick}. For these systems, it is very difficult to find
emitters which are bright enough to be detected is because both the laser
absorption and the emission rate of a radial dipole are very weak for such 
small NW diameters. We think that the emitters which have been selected
correspond to NWQDs having a large fraction of axial dipole character as they
are the brightest ones when no oxide shell is present.

\subsection{Radiation pattern}

\begin{figure}
	\centering
	\includegraphics[width =8.6cm]{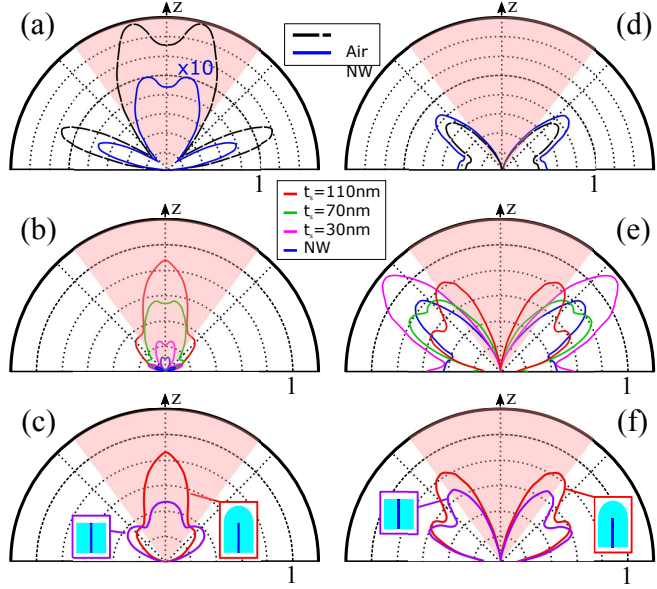}
	\caption{
		Radiation patterns from numerical simulations for a radial dipole placed at
\unit{470}{\nano\meter} above the substrate. The experimental NA region is
shaded and indicated in red. (a,d) Comparison between the case of a
free-standing emitter in air (black dashes), and embedded inside the NW (blue
solid line) for a radial (a) or axial (d) dipole.
They evidence the effect of the dielectric screening from the NW on the radial
dipole and the absence of screening for the axial dipole. (b, e)
Comparison of the total emitted intensity versus 
$t_s$ for a radial (b) or axial (e) dipole. A
combined effect of reduced dielectric screening and light guiding and
redirection towards small angles is observed. (c, f) Effect of the
shell layer termination shape for a radial (c) or axial
(f) dipole. The hemispherical shape increases the fraction of light
that is redirected towards the $z$ direction.
	}
	\label{fig:FarField}
\end{figure}

To analyze the mechanisms leading to the increase in collected intensity with
increasing shell thickness, we present in Figure~\ref{fig:FarField} several
simulated radiation patterns. They are represented as polar plots of the
far-field intensity $I(\theta)$ in the top $(x,z)$ plane, $\theta$ is the angle
between the direction of observation and the vertical $z$ axis. Simulations are
made using respectively a radial dipole [along $x$,
Figures~\ref{fig:FarField}(a-c)] or an axial dipole [along $z$,
Figures~\ref{fig:FarField}(d-f)].

Figures~\ref{fig:FarField}(a,d) show the effect of the NW structure alone (no
oxide shell being present) on such dipoles by comparing it to the case of a free
standing dipole in vacuum above the same ZnSe substrate. One can see that the
presence of the NW does not affect the shape of radiation diagram, which is
essentially determined by the interferences between the directly radiated field
and its reflection on the substrate. Most remarkably, in the case of the radial
dipole the presence of the NW dramatically reduces the emission intensity
through the dielectric screening effect discussed earlier. Simulations show a
radiative rate reduction by a factor $\sim 1/16\simeq n_{\rm ZnSe}/45$ in
agreement with the dielectric screening value predicted by
Eq.~\eqref{eq:screening}. In contrast, in the case of the axial dipole it can be
seen that the presence of the NW only slightly increases the emitted intensity.

Figures~\ref{fig:FarField}(b,e) show the computed radiation patterns of the NWQD
for increasing oxide shell thickness $t_s$. In the case of the radial dipole,
the shell first reduces the index contrast between the NW and the surrounding
medium (cf. Eq.~\ref{eq:screening}), resulting in a strong reduction of the
emitter lifetime and thus in an increased total emitted intensity as seen in in
Fig.~\ref{fig:IVsThick}(c) and Fig.~\ref{fig:TRPLVsThick}. Note that the the intensity pattern 
shown in the polar plot must be multiplied by the solid angle $\sin\theta \mathrm{d}\theta$ if one 
wants to evaluate the power radiated in the numerical aperture. This is why the intensity for an 
axial dipole can be larger than for a radial one, as seen in Fig.~\ref{fig:IVsThick}(c).
Second, as shown in Figure  \ref{fig:Semianalytical}(c), the shell presence ensures preferential
emission into the guided HE$_{11}$ mode for increasing shell thickness. As a
consequence a near-Gaussian far-field emission pattern  corresponding to the
far-field emission profile of the HE$_{11}$
mode\cite{GregersenMoerk_Controllingemissionprofile_08} is observed for
$t_s$=\unit{110}{\nano\meter}, contrary to the structures with a smaller oxide
shell thickness where one observes the presence of two closely-spaced lobes at
small emission angles ($\pm\unit{10}{\degree}$ with respect to the $z$-axis).
The resulting emission into the 0.6 NA cone is maximum for
$t_s=$\unit{110}{\nano\meter}, where the emission into the HE$_{11}$ mode is
nearly maximum [cf. Fig. \ref{fig:Semianalytical}(b)]. The effect of the oxide
shell thickness on the axial dipole is completely different. While the
total emitted intensity does not vary much, and hence the emitter lifetime stays
constant (as noted in Fig.~\ref{fig:TRPLVsThick}),  the light emitted by the
axial dipole does not couple to the HE$_{11}$ mode but is emitted exclusively
into radiation modes. Thus the fraction of intensity emitted towards the
collection lens increases only slightly as the oxide shell thickness increases
[cf. Fig.\ref{fig:FarField}(e)]. This intensity increase for the axial
dipole also presented in Fig.\ref{fig:IVsThick}(c) is not due to a change in the
spontaneous emission rate of the emitter, but rather to a slight redirection of
the emitted light.

Finally, Figures~\ref{fig:FarField}(c,f )compare the actual hemispherical geometry
of the oxide shell termination to the flat end of a simple lateral
shell. They show that the presence of the hemisphere is beneficial to the
radiation pattern for both kinds of dipole. For the radial dipole, the
hemisphere enables a near-adiabatic expansion of the HE$_{11}$
mode\cite{GregersenMoerk_Controllingemissionprofile_08} leading to a narrowing
of the far-field emission pattern and an increased collection by the numerical
aperture. The axial dipole benefits less from the hemispherical termination of
the photonic structure since no light from this dipole is coupled to the HE11
mode. We also note that half of the emitted light propagates towards the growth
substrate and due to the index-matching condition between the NW and the
substrate, this light is predominantly lost.

In order to assess the performances of our device we compute the ratio $\eta$
between the power radiated into a 0.6 NA to the one radiated into the top air
side hemisphere. This parameter is a good figure of merit for the antenna
redirection effect although it cannot be directly related to the overall
collection efficiency because of the power lost in the substrate. For our full
photonic structure and a radial dipole one has $\eta\simeq$80\% for
$t_s$=\unit{110}{\nano\meter}. This value reduces to $\simeq$66\% for a flat
terminated core-shell photonic wire illustrating the importance of the adiabatic
expansion of the HE$_{11}$ guided mode at the end of the wire. For the dipole in
the NW without shell $\eta\simeq$55\%. We have also simulated a structure
inspired by state-of-the-art devices  fabricated by top-down methods in
Ref.~\cite{Claudon2010}. In this case we simulate a \unit{110}{\nano\meter}
oxide shell photonic wire where the hemispherical termination is replaced by a
conical tapper of \ce{Al2O3} whose radius progressively decreases from 120 to
\unit{10}{\nano\meter} in \unit{1.5}{\micro\meter}. In this case one has
$\eta\simeq$94\%, showing that although beneficial our hemispherical termination
is not optimal.


\section{Conclusion}
In summary, we have presented a bottom-up approach to fabricate a dielectric
antenna around a QD inserted inside a NW. This method allows for both
reproducible and very precise fabrication of the structure on a large ensemble
of emitters at once. It is based on the deposition of a thick oxide shell around
the NW using atomic layer deposition. Experiments show a  4-fold
enhancement of the QD photoluminescence shown in Fig.~\ref{fig:IVsThick}(c)
between a \unit{20}{\nano\meter} and a \unit{110}{\nano\meter} thick shell.
Semi-analytical calculations and numerical simulations of the structure reveal that
the oxide shell thickness strongly acts on the radial dipole emission through
two main phenomena: the reduction of the dielectric screening, which increases
the spontaneous emission rate from the QD, and the redirection of light through
a waveguiding effect. Simulations suggest that the collected intensity is
multiplied by a factor 7 with respect to the bare NW case. The fabrication
process of the photonic shell is very simple and can be applied to QDs emitting
single photons up to room temperature. Although not optimal, the resulting
structure is a step towards the best nanowire single photon sources operating at
low temperature\cite{Claudon2010}. Dielectric screening could be further reduced
by growing an oxide shell of higher index matching $n_{\rm ZnSe}$ like
\ce{TiO2}. We note also that in our system a large fraction of the emitted power
is radiated in the substrate. This loss channel could be reduced by having a
mirror at the bottom of the
structure.\cite{FriedlerRobert-Philip_Efficientphotonicmirrors_08,Reimer2012}
Moreover, to fully benefit from the waveguiding approach, a better control on
the intrinsic QD properties has to be reached to ensure the presence of radial
dipoles, which radiate more efficiently in the experimental collection aperture.

\begin{acknowledgments}
This work was supported by the French National Research Agency under the contract ANR-10-LABX-51-01 and the Danish Research Council for Technology and Production (LOQIT Sapere Aude grant DFF \#4005-00370).
\end{acknowledgments}

%

\end{document}